\documentclass[twocolumn,showpacs,preprintnumbers,amsmath,amssymb]{revtex4}

\usepackage{graphicx}
\usepackage{dcolumn}
\usepackage{bm}

\begin{document}

\noindent

\preprint{}

\title{The most probable wave function of a single free moving particle}

\author{Agung Budiyono}

\affiliation{Institute for the Physical and Chemical Research, RIKEN, 2-1 Hirosawa, Wako-shi, Saitama 351-0198, Japan}

\date{\today}

\begin{abstract} 

We develop the most probable wave functions for a single free quantum particle given its momentum and energy by imposing its quantum probability density to maximize Shannon information entropy. We show that there is a class of solutions in which the quantum probability density is self-trapped with finite-size spatial support, uniformly moving hence keeping its form unchanged. 

\end{abstract}

\maketitle

{\it Introduction: Madelung Fluid} --- Since its final formulation in term of Schr\"odinger wave mechanics, quantum mechanics has claimed to have never failed any conceivable experimental test \cite{Peres book}. Yet, despite this overwhelming fact, quantum mechanics could not hide its embarrassing face that it can not give an unambiguous answer to a very simple yet fundamental question: {\it what is the wave function of a single free particle moving linearly (translationally) with finite velocity}. To describe a single free particle, one can not simply use the stationary plane wave for the reason that it extends to the whole space so that formally unnormalizable, besides is physically in contrast to our observation that a particle is spatially very well localized. In his correspondence to Einstein, Born suggested to use coherent wave packet \cite{Einstein-Born correspondence}, motivated by the fact that a coherent state is localized both in momentum and position space. Yet, coherent wave packet suffers from the lack of being not stationary to be evolved by Schr\"odinger equation. This is again is in contrast to our intuition about a particle which should have a stable structure, if left unperturbed. 

On the other hand, in our previous work \cite{AgungPRA1}, working with two dimensional Madelung fluid dynamics \cite{Madelung} whose irrotational motion reduces into the Schr\"odinger equation for a single free particle, we have developed a class of the most probable wave functions for a single free particle with a given quantum energy. This is done by imposing its quantum probability density to maximize Shannon information entropy. We showed that there is a class of solutions in which the wave function is self-trapped with finite-size spatial support, spinning around its center yet stationary. In this present paper, we shall develop yet another class of stationary solutions in which the wave function is still self-trapped with finite-size support and moving linearly with uniform (thus definite) velocity. We then end to give an interpretation to the wave function as describing a single free particle moving with finite and definite velocity. 

\begin{figure}[tbp]
\begin{center}
\includegraphics*[width=7cm]{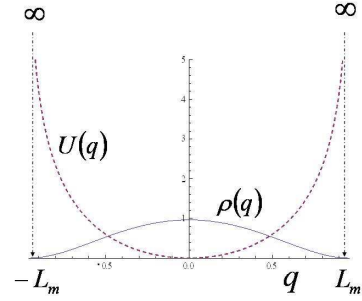}
\end{center}
\caption{The profile of a self-trapped quantum probability density $\rho(q)$ (solid line) and the corresponding quantum potential $U(q)$ (dashed line) that it generates. The quantum potential is shifted down so that its global minimum is equal to zero. See text for detail.}
\label{1D QP-QPD}
\end{figure}

Let us consider the Madelung fluid dynamics of a single free particle with mass $m$.  Since in the present paper we are only interested in the translational motion of the particle, it is then sufficient to consider a spatially one dimensional case. First, the state of Madelung fluid in configuration space $q$ is determined by a pair of fields, $\{\rho(q),v(q)\}$. $\rho(q)$ is called as quantum probability density thus assumed to be always normalized. Whereas $v(q)$ is a velocity vector field. Their evolution in time $t$ is then governed by the following coupled differential  equations:   
\begin{eqnarray}
m\frac{dv}{dt}=-\partial_qU,\hspace{2mm}\partial_t\rho+\partial_q(v\rho)=0.
\label{Madelung fluid}
\end{eqnarray}
Here, $U(q)$ is called as quantum potential determined by the quantum probability density as 
\begin{equation}
U(q)= -\frac{\hbar^2}{2m}\frac{\partial_q^2R}{R},
\label{quantum potential}
\end{equation} 
where $R(q)=\rho^{1/2}(q)$ is quantum amplitude. Notice that for one dimensional system, one can always derive the velocity field from a scalar field $S(q)$ as $v(q)=\partial_qS/m$. Using this to define a complex-valued wave function as $\psi(q)=R(q)\exp(iS(q)/\hbar)$, Madelung fluid dynamics of equation (\ref{Madelung fluid}) is then equal to the Schr\"odinger equation for a single free particle:
\begin{equation}
i\hbar\partial_t\psi(q;t)=-\frac{\hbar^2}{2m}\partial_q^2\psi(q;t).
\label{Schroedinger equation}
\end{equation}
In this regard, $S(q)$ is called as quantum phase. 

{\it The most probable--self-trapped quantum probability density} --- Next, let us consider a class of quantum probability densities which maximizes Shannon information entropy: $H[\rho]=-\int dq\hspace{1mm}\rho(q)\ln\rho(q)$ \cite{Shannon entropy},  given its average quantum potential $\bar{U}=\int dq\hspace{1mm}U(q)\rho(q)$. This is the so-called maximum entropy principle \cite{Jaynes-MEP}. It has been argued as the only way to infer from an incomplete information which does not lead to logical inconsistency \cite{Shore-Johnson-MEP}. Hence, it will give us the most probable quantum probability density with average quantum potential $\bar{U}$. This inference  problem can be directly solved to give \cite{Mackey-MEP}:
\begin{equation}
\rho(q)=\frac{1}{Z}\exp\big(-U(q)/T\big),
\label{canonical QPD}
\end{equation}
where $T$ is the Lagrange constant below chosen to be non-negative, and $Z(T)$ is a normalization factor. Combining with equation (\ref{quantum potential}), equation (\ref{canonical QPD}) comprises a differential equation for $\rho(q)$ or $U(q)$, subjected to the condition that $\rho(q)$ must be normalized. In term of $U(q)$, one has \cite{AgungPRA1}
\begin{equation}
\partial_q^2U=\frac{1}{2T}(\partial_qU)^2+\frac{4mT}{\hbar^2}U. 
\label{NPDE for U}
\end{equation}

Figure \ref{1D QP-QPD} shows the solution of equation (\ref{NPDE for U}) with the boundary conditions: $U(0)=1$, $\partial_qU(0)=0$, for $T=1$. All numerical results in this paper are obtained by setting $\hbar=m=1$. One can see that the quantum probability density is being trapped by its own self-generated quantum potential. The self-trapping property is valid for any positive value of $U(0)$ provided that $\partial_qU(0)=0$. This can be understood as follows. First, at the origin, $q=0$, we have $\partial_q^2U(0)=(4mT/\hbar^2)U(0)>0$, hence $U(q)$ is convex at $q=0$. Since $\partial_qU(0)=0$, then at points nearby $q=0$, one concludes that $U(q)$ is positive definite. Furthermore, since the first term on the right hand side of equation (\ref{NPDE for U}) is always non-negative, then at these points one has $\partial_q^2U>0$. This reasoning can be extended to the whole space so that $U(q)$ is everywhere convex and positive definite. Using this fact in equation (\ref{canonical QPD}), one concludes that $\rho(q)$ is trapped by its own $U(q)$.  

Next, from the definite positivity of $U(q)$, then the definition of quantum potential of equation (\ref{quantum potential}) gives us  
\begin{equation}
\partial_q^2R<0. 
\label{concave quantum amplitude}
\end{equation}
$R(q)$ is thus concave everywhere. Since $R(q)$ is finite and possesses symmetry $R(-q)=R(q)$, then $R(q)$ must cross the $q-$axis at finite values of $q=\pm L_m$ where $U(\pm L_m)=\infty$. See Fig. \ref{1D QP-QPD}. Hence, the self-trapped quantum probability density developed in previous paragraph possesses only a finite range of spatial support given by the interval: $\mathcal{M}=[-L_m,L_m]$. 

\begin{figure}[tbp]
\begin{center}
\includegraphics*[width=7cm]{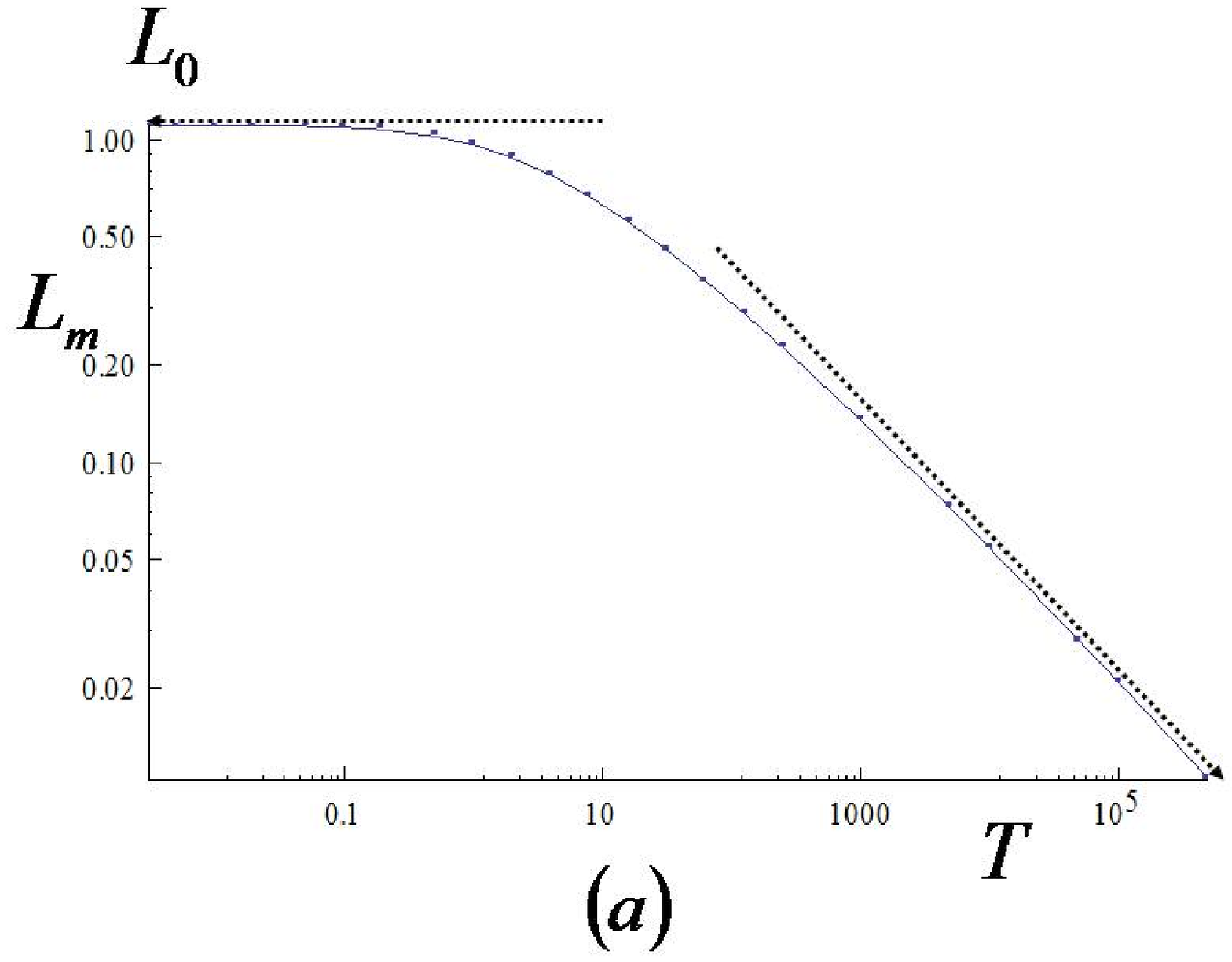}
\includegraphics*[width=7cm]{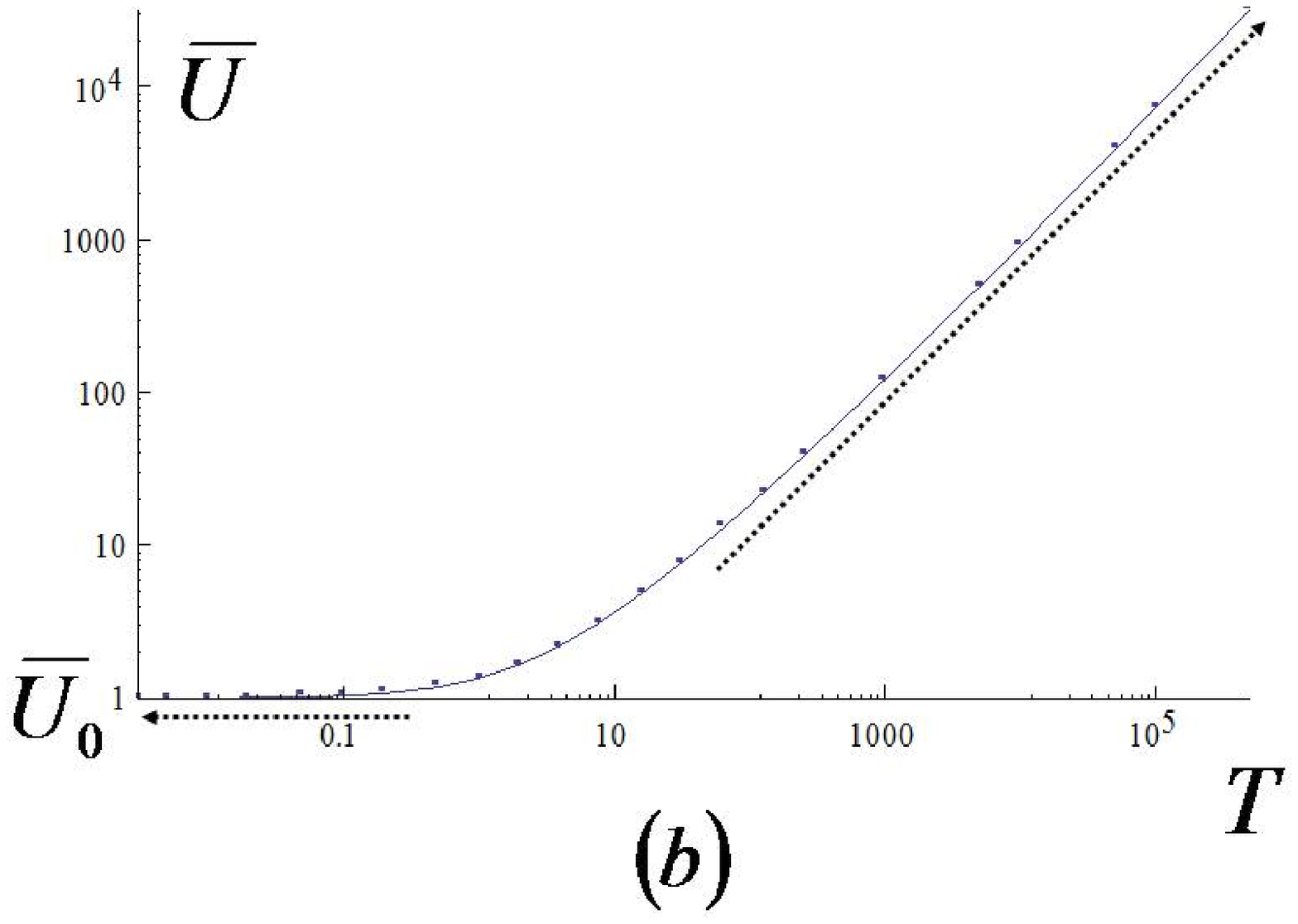}
\end{center}
\caption{(a) Half length $L_m$ of the support of self-trapped quantum probability  density, and (b) its average quantum potential $\bar{U}$ plotted against $T$.}
\label{1D temperature vs rmax}
\end{figure}

In Fig. \ref{1D temperature vs rmax}, we plot (a) $L_m$, and (b) the average of the quantum potential $\bar{U}=\int dqU\rho$, of our self-trapped quantum probability density against the variation of the parameter $T$, by solving equation (\ref{NPDE for U}) with fixed boundary conditions: $\partial_qU(0)=0$ and $U(0)=1$. Two asymptotic limits are of great interest. First,  for the case when $T$ takes an infinite value, $T\rightarrow\infty$, $L_m$ is approaching zero, $\lim_{T\rightarrow \infty}L_m=0$. One thus concludes that the quantum probability density is approaching a delta function for sufficiently large value of $T$. Further, one can also see that in the limit $T\rightarrow\infty$, the average quantum potential is infinite,  $\lim_{T\rightarrow\infty}\bar{U}\rightarrow\infty$. Based on this last fact, we shall show later that the limit $T\rightarrow\infty$ is not relevant physically. 

\begin{figure}[tbp]
\begin{center}
\includegraphics*[width=7cm]{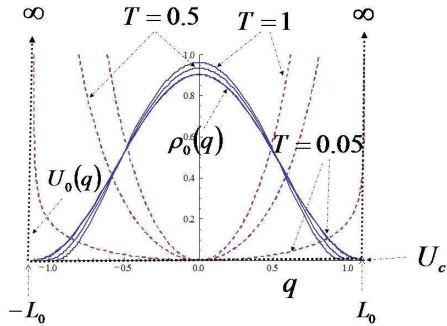}
\end{center}
\caption{The quantum probability density (solid line) and its corresponding quantum potential (dashed line) for several small values of $T$ obtained by solving equation (\ref{NPDE for U}). We also plot the analytical solution for $\rho_0(q)$ at $T=0$, assuming that $U_0(q)$ takes the form of a box with infinite wall. See text for detail.}
\label{small temperature vs QPD and QP}
\end{figure}

On the contrary, for vanishing value of $T$, we see from Fig. \ref{1D temperature vs rmax} that both $L_m$ and $\bar{U}$ are approaching some finite values, $\lim_{T\rightarrow 0}L_m=L_0$ and $\lim_{T\rightarrow 0}\bar{U}=\bar{U}_0$. This hints us that the quantum probability density and thus its corresponding quantum potential are also converging toward some functions
\begin{eqnarray}
\lim_{T\rightarrow 0}U(q;T)=U_0(q),\hspace{2mm}\lim_{T\rightarrow 0}\rho(q;T)=\rho_0(q).
\label{QPD and QP at vanishing temperature}
\end{eqnarray}
Let us discuss this asymptotic situation in more detail. To do this, in Fig. \ref{small temperature vs QPD and QP} we plot the profile of quantum probability density and its corresponding quantum potential for several small values of $T$, $T=1,0.5,0.05$, by solving equation (\ref{NPDE for U}) with fixed boundary conditions: $\partial_qU(0)=0$ and $U(0)=1$.  Let us first pay our attention to the profile of quantum potential $U(q;T)$. One can see that as $T$ decreases, the quantum potential inside the support is getting flatterer before becoming infinite at the boundary points, $q=\pm L_m(T)$. One can thus guess that at vanishing $T$, $T=0$, the quantum potential is converging toward a box of length $2L_0$ with perfectly flat bottom and an infinite wall at the boundary points, $q=\pm L_0$. See Fig. \ref{small temperature vs QPD and QP}. 

Based on the above guess, let us proceed to calculate the profile of the quantum probability density for vanishing $T$, $\rho_0(q)$. To do this, let us denote the assumed constant value of quantum potential as $U_0(q)=U_c$, for $-L_0<q<L_0$. Since $U_0(q)$ is flat inside the support, then one has $\bar{U}_0=\int_{-L_0}^{L_0} dq\rho_0(q)U_0(q)=U_c$. Hence, from the definition of quantum potential given in equation (\ref{quantum potential}), inside the support of the quantum probability density, $\mathcal{M}$, one has 
\begin{equation}
-\frac{\hbar^2}{2m}\partial_q^2R_0=U_cR_0=\bar{U}_0R_0,
\label{stationary Schroedinger equation 1}
\end{equation} 
where $R_0(q)\equiv\rho_0^{1/2}(q)$ is the quantum amplitude at $T=0$. The above differential equation must be subjected to the boundary condition: $R_0(\pm L_0)=0$. 

Next, solving equation (\ref{stationary Schroedinger equation 1}) one obtains
\begin{equation}
R_0(q)=A_0\cos(k_0 q),
\label{localized-stationary QPD}
\end{equation}
where $A_0$ is a normalization constant and $k_0$ is related to the average quantum potential as 
\begin{equation}
k_0=\sqrt{2m\bar{U}_0/\hbar^2}. 
\label{wave number vs average quantum potential}
\end{equation}
The boundary condition imposes $k_0L_0=\pi/2$. In Fig. \ref{small temperature vs QPD and QP}, we plot the above obtained quantum probability density, $\rho_0(q)$. One can see that as $T$ is decreasing toward zero, $\rho(q;T)$ obtained by solving the differential equation of (\ref{NPDE for U}) is indeed converging toward $\rho_0(q)$ given in equation (\ref{localized-stationary QPD}). This confirms our initial guess that at $T=0$ the quantum potential takes a form of box with infinite wall at $q=\pm L_0$. 

{\it Stationary-moving wave packet} --- Let us now choose the pair of fields $\{\rho_0(q),v_0(q)\}$ as the initial state of the Madelung fluid. Here, $\rho_0(q)=R_0^2(q)$ is given in equation (\ref{localized-stationary QPD}) and $v_0(q)=v_c$ is a uniform velocity field with non-vanishing value only inside the spatial support $\mathcal{M}$. Then, since at $t=0$ the quantum potential is flat inside the support, the quantum force is initially vanishing, $\partial_qU=0$. Hence, from the dynamical equation (\ref{Madelung fluid}), the velocity field at infinitesimal lapse of time $t=\Delta t$ will stay unchanged and keeps uniform. This in turn will shift the quantum probability density in space by $\Delta q=v_c\Delta t$ while keeps its spatial profile unchanged. Correspondingly, it will shift the support of the quantum probability density by the same amount: $\mathcal{M}\rightarrow \mathcal{M}_{\Delta t}=[v_c\Delta t-L_0,v_c\Delta t+L_0]$. The same thing occurs for the next infinitesimal time lapse and so on and so forth. Hence, at finite time lapse, the velocity field is kept uniform with constant value. This fact finally gives us
\begin{equation}
d_t\rho=\partial_t\rho+v_0\partial_q\rho=\partial_t\rho+\partial_q(v_0\rho)=0, 
\label{stationary quantum probability density}
\end{equation}
where in the second equality we have used the continuity equation of (\ref{Madelung fluid}) and the fact that the velocity field is uniform so that $\partial_qv_0=0$. One thus concludes that \textit{the quantum probability density is moving with a uniform velocity field $v_c$, keeping its initial form remained unchanged}. 

Hence, at time $t$, the quantum probability density is given by 
\begin{eqnarray}
\rho(q;t)=\rho_0(q-v_ct)=A_0^2\cos^2(k_0q-\omega_0 t),
\label{stationary-moving quantum probability density}
\end{eqnarray}
where $\omega_0=k_0v_c$ and $q\in [v_ct-L_0,v_ct+L_0]$. One can check easily that equation (\ref{stationary-moving quantum probability density}) indeed satisfies the continuity equation of (\ref{Madelung fluid}). Moreover, the quantum phase at time $t$ can be obtained by integrating $v(q)=(1/m)\partial_qS=v_c$ to give
\begin{equation}
S(q;t)=mv_cq+\xi(t), 
\end{equation}
where $\xi(t)$ is a function which depends at most only on $t$. We shall give the explicit form of $\xi(t)$ later thereby revealing its physical meaning. The wave function we are searching thus takes the following form: 
\begin{equation}
\psi(q;t)=A_0\cos\big(k_0(q-v_ct)\big)\exp\Big(\frac{i}{\hbar}\big(mv_cq+\xi(t)\big)\Big),
\label{stationary-moving wave function}
\end{equation} 
where $q\in [v_ct-L_0,v_c+L_0]$.

Next, let us calculate the quantum mechanical energy of our  self-trapped, uniformly moving-stationary wave function. Since quantum mechanical energy is conserved then it is sufficient to use the wave function at $t=0$. Putting the wave function in polar form, $\psi(q)=R_0(q)\exp(iS(q)/\hbar)$, one has 
\begin{eqnarray}
\langle E\rangle\equiv \int_{-L_0}^{L_0} dq\hspace{1mm}\psi^*(q)\Big(-\frac{\hbar^2}{2m}\partial_q^2\Big)\psi(q)\hspace{10mm}\nonumber\\
=\int_{-L_0}^{L_0} dq\hspace{1mm}\Big(-\frac{\hbar^2}{2m}R_0\partial_q^2R_0+\frac{1}{2m}R_0^2(\partial_qS)^2
\nonumber\\-\frac{i\hbar}{m}R_0\partial_qR_0\partial_qS-\frac{i\hbar}{2m}R_0^2\partial_q^2S \Big).
\label{calculation of quantum mechanical energy}
\end{eqnarray} 
The first term on the right hand side is equal to the average quantum potential, $\bar{U}_0=\int dqU_0\rho_0=\hbar^2k_0^2/(2m)$. Next, defining kinetic energy density as $K_0(q)\equiv(m/2)v_0^2(q)$, then the second term is equal to the kinetic energy of the Madelung fluid: 
\begin{equation}
\bar{K}_0=\int dq K_0(q)\rho_0(q)=\frac{mv_c^2}{2}. 
\label{kinetic energy of Madelung fluid}
\end{equation}
Further, for a uniform velocity field, the last term is vanishing, $(1/m)\partial_q^2S=\partial_qv_0=0$. Again for a uniform velocity field, since $R_0(q)$ is an even function and $\partial_qR_0(q)$ is an odd function then the third term is also vanishing. Hence, in total, the quantum mechanical energy of the  self-trapped wave function moving with a uniform velocity field can be decomposed as 
\begin{equation}
\langle E\rangle=\bar{U}_0+\bar{K}_0=\frac{\hbar^2k_0^2}{2m}+\frac{mv_c^2}{2}.
\label{energy decomposition}
\end{equation}
Moreover, using similar argument as above, the average quantum momentum can also be calculated to give
\begin{eqnarray}
\langle p\rangle\equiv\int_{-L_0}^{L_0}dq\hspace{1mm}\psi^*(q)\big(-i\hbar\partial_q\big)\psi(q)=mv_c. 
\label{average momentum}
\end{eqnarray}
The average kinetic energy of the Madelung fluid and average quantum momentum are thus related as $\bar{K}_0=\langle p\rangle^2/(2m)$. This observation leads us to conclude that $\bar{U}_0$ must essentially be interpreted as an {\it internal energy} of the single particle. It is this energy that is missed to be taken into account if one uses the simple plane wave to represent a moving particle. 

Finally, putting the wave function of the form (\ref{stationary-moving wave function}) into the Schr\"odinger equation of (\ref{Schroedinger equation}), keeping in mind  $\omega_0=k_0v_0$ and equation (\ref{energy decomposition}), one easily obtains  $\partial_t\xi=-\langle E\rangle$, which can be integrated to give 
\begin{equation}
\xi(t)=-\langle E\rangle\hspace{1mm} t, 
\label{energy vs phase}
\end{equation}
modulo to some constant. Putting everything into equation (\ref{stationary-moving wave function}), one finally obtains
\begin{equation}
\psi(q;t)=A_0\cos\Big(k_0\Big(q-\frac{\langle p\rangle}{m}t\Big)\Big)\exp\Big(\frac{i}{\hbar}\big(\langle p\rangle q-\langle E\rangle t\big)\Big),
\label{stationary-moving wave function final}
\end{equation}
where $q\in [\langle p\rangle t/m-L_0,\langle p\rangle t/m+L_0]$. Hence, since $k_0$ is determined by fixing the values of $\langle E\rangle$ and $\langle p\rangle$, the stationary-moving wave function we developed in this paper can be interpreted as describing {\it the most probable wave functions for a single free moving particle with momentum $\langle p\rangle=mv_c$ and total energy $\langle E\rangle$}. 

{\it Conclusion and Discussion} --- To conclude, we have developed a class of the most probable wave functions for a single quantum particle moving with finite velocity. This is done by imposing the wave function to maximize Shannon information entropy. We showed that there is a class of solutions in which the wave function is self-trapped by its own quantum potential such that it is localized with finite-size spatial support; translationally moving with uniform velocity thus keeping its form unchanged. We also showed that there is a new type of energy, referred to as internal energy, which is missed if one uses a plane wave to represent the free moving particle.  

The fact that the velocity field is uniform in space asserts that the momentum of the wave packet is {\it definite}. Surprisingly, even in this case, the wave packet possesses only a {\it finite uncertainty} in position. This observation thus raises an interesting question concerning the physical status of Heisenberg uncertainty principle. 

\begin{acknowledgments} 

This research is funded by the FPR program in RIKEN. The author acknowledges useful discussion with Masashi Tachikawa and Ken Umeno. 

\end{acknowledgments}

\end{document}